\documentclass[12pt,english]{iopart}
\usepackage[T1]{fontenc}
\usepackage[latin9]{inputenc}
\usepackage{amssymb}
\usepackage{graphicx}
\usepackage{subscript}

\makeatletter
\@ifundefined{textcolor}{}
{%
 \definecolor{BLACK}{gray}{0}
 \definecolor{WHITE}{gray}{1}
 \definecolor{RED}{rgb}{1,0,0}
 \definecolor{GREEN}{rgb}{0,1,0}
 \definecolor{BLUE}{rgb}{0,0,1}
 \definecolor{CYAN}{cmyk}{1,0,0,0}
 \definecolor{MAGENTA}{cmyk}{0,1,0,0}
 \definecolor{YELLOW}{cmyk}{0,0,1,0}
}

\makeatother

\usepackage{babel}

\newcommand{\JkgK}{Jkg$^{-1}$K$^{-1}$}

\begin{document}

\title{Electronic structure, metamagnetism and thermopower of LaSiFe$_{12}$ and interstitially doped LaSiFe$_{12}$}

\author{Z. Gercsi}

\address{CRANN and School of Physics, Trinity College Dublin, Dublin 2, Ireland}
\address{Dept. of Physics, Blackett Laboratory, Imperial College London, London
SW7 2AZ, United Kingdom}

\author{N. Fuller}
\address{Dept. of Physics, Brooklyn College, CUNY, 2900 Bedford Avenue, Brooklyn, New York 11210, USA}

\author{K.G. Sandeman}
\address{Dept. of Physics, Blackett Laboratory, Imperial College London, London
SW7 2AZ, United Kingdom}
\address{Dept. of Physics, Brooklyn College, CUNY, 2900 Bedford Avenue, Brooklyn, New York 11210, USA}
\address{Physics Program, The Graduate Center, CUNY, 365 Fifth Avenue, New York, New York 10016, USA}

\author{A. Fujita}

\address{Green-Innovative Magnetic Material Research Center, National Institute
of Advanced Industrial Science and Technology Anagahora 2266-98, Simosidami,
Nagoya 463-8560, Japan }

\begin{abstract}
We present a systematic investigation of the effect of H, B, C, and N interstitials on the electronic, lattice and magnetic properties
of La(Fe,Si)$_{13}$ using density functional theory.   The parent LaSiFe$_{12}$ alloy has a shallow, double-well free energy function that is the basis of itinerant metamagnetism.  On increasing the dopant concentration, the resulting lattice expansion causes an initial increase in magnetisation for all interstitials that is only maintained at higher levels of doping in the case of hydrogen.  Strong \emph{s-p} band hybridisation occurs at high B,C and N concentrations.  We thus find that the electronic effects of hydrogen doping are much less pronounced than those of other interstitials, and result in the double-well structure of the free energy function being least sensitive to the amount of hydrogen.  This microscopic picture accounts for the vanishing first order nature of the transition by B,C, and N dopants as observed experimentally.    We use our calculated electronic density of states for LaSiFe$_{12}$ and the hydrogenated alloy to infer changes in magneto-elastic coupling and in phonon entropy on heating through $T_C$ by calculating the fermionic entropy due to the itinerant electrons.  Lastly, we predict the electron thermopower in a spin-mixing, high temperature limit and compare our findings to recent literature data.
\end{abstract}
%\submitto{JPD}
\maketitle
\ioptwocol

\section{Introduction}

The magnetocaloric effect (MCE) is the temperature change
of a substance subjected to a change in applied magnetic field. The discovery of the effect can  be attributed to Weiss and Piccard's observation of the magnetization of nickel close to its Curie point in 1917~\cite{WeissPicc}, after a recent re-examination of the original literature by Smith~\cite{Anders}. The adiabatic demagnetization of paramagnetic salts was shown by Giauque and MacDougall in 1933\cite{GiaMac1933} following initial proposals by both Debye and Giauque in the previous decade~\cite{Debye,Giauque}.  The study of room temperature MCEs associated with a magnetic phase transition was revived in 1997 by Pecharsky and Gschneidner who observed a `giant' entropy change of $\sim$14~JK$^{-1}$kg$^{-1}$ in a 0-2~Tesla field change in Gd$_{5}$Si$_{2}$Ge$_{2}$~\cite{Gd5(Si2Ge2)}.  While experiments up to that point had postulated the possibility of room temperature refrigeration using, for example, a second order Curie transition such as that found in Gd~\cite{brown_1976}, it was the effects associated with the first order transition seen in Gd$_{5}$Si$_{2}$Ge$_{2}$ that initiated widespread research interest in the MCE. Today, a large set of magnetic materials show `large' or `giant' magnetocaloric effects~\cite{SandemanScripta} and form one family in a more general set of ferroic refrigerants~\cite{takeuchi_2015}.  However, a good refrigerant material also needs to fulfil auxiliary requirements such as tuneable thermal conductivity, durability and elemental abundance and so the number of material systems that are close to commercialisation is relatively small.  This situation provides motivation for the use of theoretical models that may aid the understanding and prediction of caloric effects.

Density functional theory (DFT) is a valuable tool with which to describe the changes in matter at the electronic level that may lead to a large MCE.   Elemental Gd has a ferromagnetic (FM) ordering temperature around room temperature that makes it an ideal candidate magnetocaloric material. DFT calculations based on thermally induced spin fluctuations in a disordered local moment picture showed that the magnetic order in Gd is linked to the $c/a$ ratio and atomic unit cell volume~\cite{Gd-Nature}. Such magneto-elastic coupling is useful for generating a large MCE since it increases the rate of change of magnetization with temperature.   In Gd$_{5}$(Si$_{2}$Ge$_{2}$), DFT calculations indicated breaking and reforming of Si-Ge bonds between layers within the unit cell, affecting both the location of the Fermi level and the effective magnetic exchange coupling, increasing the latter to the level where a first order magneto-structural transition is observed~\cite{Gd5(Si2Ge2)}.  However, the cost of heavy rare-earth Gd renders magnetocaloric alloys with high d-metal content preferable~\cite{bruckreview_2005}.

Manganites and manganese silicides have also been the subject of DFT studies.  In manganites, the broad variety of crystallographic, magnetic and electronic phases are attributed to the strong interplay between spin, charge, orbital and lattice degrees of freedom that often couples to external magnetic fields and results in measurable MCE.  For a qualitative description of these correlated physical quantities, state-of-the-art hybrid exchange
density functionals can be applied~\cite{Romi_manganites}. In the case of manganese based metallic silicides, ground state and finite temperature DFT models have been used to model and predict new Mn-based metamagnets~\cite{Zsolt2,Zsolt1,TDFT_Staunton}.  Those calculations used accurate structural data obtained from high resolution neutron diffraction on CoMnSi, a noncollinear antiferromagnet (AFM) that exhibits giant magneto-elastic coupling~\cite{Alex_PRL}.

Experimentally, the most intensively studied MCE materials are based on either Fe$_{2}$P or La(Fe,Si)$_{13}$.  Both have been the subject of some modelling studies. In Fe$_{2}$P, iron has two inequivalent crystallographic sites and the low moment site (3$f$) has a metamagnetic transition\cite{Yamada} at the Curie temperature, 212~K.  The Curie point can be tuned through room temperature by partial replacement of Fe by Mn as well as P by, for example, Si~\cite{Dung1}. The so-called mixed magnetism of this material has been investigated by a number of DFT studies that have identified the mechanism of magneto-elastic coupling and the change of electron density across the Curie transition~\cite{Dung1,Erna1,Erna2,Fe2P_ZG1,Liu1}.

In this article, we perform a DFT study of compounds based on LaFe$_{13-z}$M$_{y}$ (M=Si, Al), a cubic NaZn$_{13}$-type material which was first synthesized by Kripyakevich et al.~\cite{Krypiakewytsch}.  To date, much of the compositional tuning that is used to adjust the magnetocaloric effect and its temperature range is the result of empirical work rather than theory-led prediction.  As a function of Si (Al) content, LaFe$_{13-y}$M$_{y}$ exhibits an FM (or AFM) transition upon cooling at a temperature between 180 and 250 K with large MCE. Furthermore, the magnetic field-dependent itinerant-electron metamagnetic (IEM) transition\cite{IEM1,IEM2} can be shifted towards room temperature by Si addition. However, on increasing the Si content above $y>1.8$, a change in the nature of the FM phase transition from first-order to second-order takes place that results in a considerable reduction of the useful MCE.  A first-principles calculation by Wang et al.~\cite{Wang} indicated that hybridization between the Fe-$d$ and Si-$p$ states is linked to the reduction of Fe magnetic moment as well as to the smearing of the first-order type transition for alloys with high Si-content.

The partial replacement of the transition metal element Fe by Co or Mn has been explored in an attempt to preserve the first order nature of the transition around the Curie temperature, $T_C$, although both elements cause significant weakening of the field-induced IEM transition~\cite{Liu-Co}. Similarly, interstitial doping of $s$-block or $p$-block elements has also been pursued experimentally in order to raise the IEM to room temperature. These empirical studies found that the preparation of single phase compositions is limited to low interstitial concentrations and that only hydrogen is capable of the increase of magnetic transition to room temperature without the diminution of useful isothermal entropy change~\cite{chen_2003,teixeira_2012a,ZhangCdope,zhang_2010,balli_2009a,fujita_2003a}. Theoretical calculations by Kuz\textquoteright{}min and Richter~\cite{Kuzmin_DFT} on LaFe$_{12}$Si, without interstitial substitution, found that the free energy, as a function of magnetization ($F(M)$) has several shallow minima and maxima, to which they attributed the reduced
hysteresis and improved magnetocaloric performance of La(Fe,Si)$_{13}$.  Fujita and Yako \cite{FujitaScripta} extended this approach and further detailed the dependence of such an energy plot on both the lattice size and the degree of Fe/Si substitution.  Most recently, Gruner et al. used a DFT approach to model the difference in phonon density of states between the paramagnetic and ferromagnetic states~\cite{gruner_2015}.  They found, consistent with nuclear resonant inelastic x-ray scattering (NRIXS), that phonon entropy appears on heating through $T_C$, despite the decrease in unit cell volume.

We note that no systematic investigation of the effect of interstitial $s$-block or $p$-block elements on the electronic, lattice and magnetic properties of La(Fe,Si)$_{13}$ had been carried out before this work. The work presented here, using a theoretical approach based on DFT, attempts to describe the effect of the size of four different dopants and their valence electrons to understand how interstitials can influence the magnetocaloric performance of these alloys.  We use our calculated electronic density of states to further examine two experimental quantities linked to the entropy change present at $T_C$.  Firstly, we infer changes in magneto-elastic coupling and in phonon entropy on heating through $T_C$ by calculating the fermionic entropy due to the itinerant electrons.  Secondly, we predict the electron thermopower in a spin-mixing, high temperature limit and compare our findings to recent literature data.  We describe our theoretical methods in section~\ref{sec_methods} before presenting our results and discussion in section~\ref{sec_res-dis}.  Conclusions are drawn in section~\ref{sec_conc}.

\section{Methods}
\label{sec_methods}

\subsection{Computational models used \label{sub:Theoretical-model}}

Our computational approach is divided into two, complementary parts.
In the first part, we investigate the effect of interstitial doping on the equilibrium unit cell volume using the projector augmented wave (PAW)
method~\cite{VASP} as implemented in the Vienna ab-initio simulation package (VASP). The VASP code with Perdew-Burke-Ernzerhof (PBE) parameterization~\cite{PBE} is employed, where site-based magnetic moments were calculated using the Vosko-Wilk-Nusair interpolation~\cite{Vosko} within the general gradient approximation (GGA) for the exchange-correlation potential. 

La(Fe,Si)$_{13}$ has 8 formula units per conventional cell. The La atoms occupy the 8$a$ sites ($\frac{1}{4}$,$\frac{1}{4}$,$\frac{1}{4}$),
while diffraction studies show that Fe and Si atoms can occupy both 8$b$ $(000)$
and the 96$i$ $(0yz)$ crystallographic positions~\cite{LaFe-Si1,LaFe-Si2,LaFe-Si3,LaFe-Si4}.  In order to keep the computational requirements at a feasible level, we follow the approach previously adopted by Kuz\textquoteright{}min and Richter~\cite{Kuzmin_DFT}, limiting our investigations to an atomically
ordered version of LaSiFe$_{12}$Z$_{x}$, where the 8$b$ sites are occupied solely by silicon whilst iron is located exclusively on the
96$i$ sites. In such a case, the cell that forms the basis of the calculations contains
2 La, 26 Fe and 2 Si atoms. Furthermore, interstitial elements H, B, C, and N were considered to occupy the 24$d$ crystallographic
site only. Using this model, we may vary the concentration of interstitials, $x_Z$ in steps of 0.5 from $x=0$ to 3 in LaSiFe$_{12}$Z$_{x}$.  For a picture of the crystal structure including interstitial sites, we refer the reader to Figure~1 of Fujieda et al.~\cite{fujieda_neutron_2008}.

Full structural relaxation was carried out for both collinear ferromagnetic (FM) and non-magnetic (NM) states in the case of parent LaFe$_{12}$Si alloy, while only the lattice parameter $a$ was relaxed (without relaxation of the internal atomic positions) for the materials doped with $s$- or $p$-block interstitials. A 7 $\times$ 7 $\times$ 7 $k$-point grid  was used to discretize the first Brillouin zone and the energy convergence criterion was set to $5\times10^{-7}$~eV during the energy minimization process. The effect of spin-orbit coupling was tested for the parent alloy, where we found negligible contributions to the magnetic moments ($<10^{-3}\mu_{B}$) and total energies ($<10^{-6}$~eV) and thus it was turned off for the calculations presented here. Finally, data presented in Fig.~\ref{fig:PDOS} was calculated on a dense 19 $\times$ 19 $\times$ 19 grid of $k$-points for high accuracy.

In the second part of this study, we have taken a fixed spin moment (FSM) approach within the tight-binding theorem using linear muffin tin
orbitals (TB-LMTO) as implemented in the Stuttgart TB-LMTO code~\cite{LMTO,Stuttgartcode,FSM}. This method requires carefully adjusted overlapping Wigner-Seitz (WS) atomic spheres included in the calculations to complete the basis and to provide an accurate description of the electron density throughout the entire unit cell. Consequently, the structural parameters of the relaxed lattice are inherently dependent on the volume occupied by the WS spheres and/or empty spheres. For this reason, we used VASP code (see above) for relaxation. Nevertheless, the TB-LMTO approach allows us to evaluate the total energy difference between FM and NM states, $\Delta F(M)$ as a function of fixed spin moment $M$ as well as the corresponding density of states (DOS) and band dispersions.  A dense mesh with 48$\times$48$\times$48 $k$-points (for the DOS calculations) or with 12$\times$12$\times$12 $k$-points (for the FSM calculations) was used.

\section{Results and Discussion}
\label{sec_res-dis}

\subsection{Effect of dopants on lattice expansion and magnetic properties }

\begin{figure}
\includegraphics[width=\columnwidth]{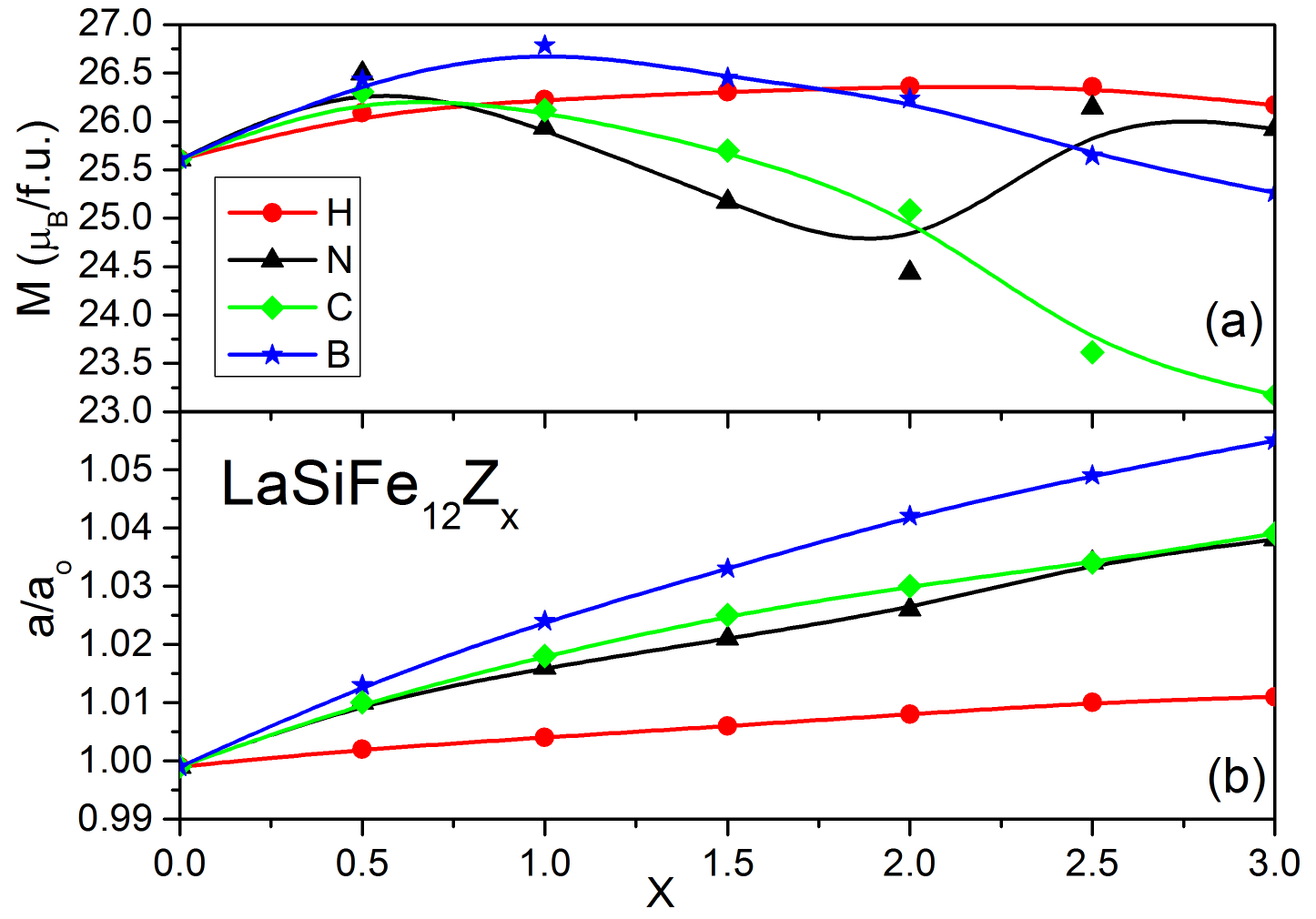}

\caption{\label{fig:a/a0} Magnetic moment/formula unit (top) and normalised lattice expansion ($a/a_{0}$)
(bottom) as a function of dopant concentration in LaSiFe$_{12}$Z$_{x}$. The lattice parameter $a_{0}$
corresponds to the fully relaxed, FM structure free of interstitials. }
 
\end{figure}
Fig. \ref{fig:a/a0} shows the calculated lattice parameter in the FM
state in LaSiFe$_{12}$Z$_{x}$ as a function of interstitial doping. Our calculations obtain a relaxed structure that differs by only about
0.1\% from the experimentally reported value. This remarkable agreement validates our choice of exchange correlation, GGA.  The lattice
expansion increases monotonically with dopant concentration at a rate that depends strongly on the size of the interstitial element.  The empirical atomic radius of hydrogen (25~pm) is much smaller than that of the boron (85~pm), carbon (70~pm), or nitrogen (65~pm).  The trend in calculated lattice expansion in Fig.~\ref{fig:a/a0} correlates well with the relative atomic size of the interstitial, showing the predominant influence of the latter on the size of the unit cell.  Our calculations are consistent with the experimental values for X=H and C interstitials respectively~\cite{Jia2_Hdoping,ZhangCdope}.  At full doping, we here find a relative lattice expansion of 0.4\% for hydrogen and a considerably higher value of 1.7\% for carbon. The values of $\frac{\triangle a}{a}$ are 1.8\% and 1.25\% for Z=B and N respectively at full doping, but these are yet to be confirmed experimentally.  There is only limited experimental data available on the relative effects of lattice expansion of the dopants studied here, particularly as full occupation of the 24$d$ site ($x=3$) by any of the dopants has not been achieved in practice.  

In terms of valence electron number, however, a different sequence exists: H$(s{}^{1})=$ B$(p{}^{1})<$ C$(p{}^{2})<$ N$(p{}^{3}$).  A closer
look at  Fig.~\ref{fig:a/a0} reveals a non-monotonic behaviour in the magnetic moment per formula unit as a function of doping, especially in the case of nitrogen.  Indeed, additional charges significantly alter the electronic structure (apart from H) in ways that go beyond the simple picture of chemical pressure effects, as we discuss later.  The contribution of additional valence electrons is also reflected in the calculated magnetic moment ($M$). $M$ rises initially (Fig. \ref{fig:a/a0}, top) for each dopant but a monotonic increase up to $x=2.5$ is only seen in the case of hydrogen.  These observations imply a mechanism
for hybrid band formation and band broadening for any dopant with a larger atomic radius than that of hydrogen. In order to depict the changes
in the electronic structure that are brought about by the interstitial elements, we next examine the partial
electronic density of states (PDOS). 

\begin{figure}
\includegraphics[width=\columnwidth]{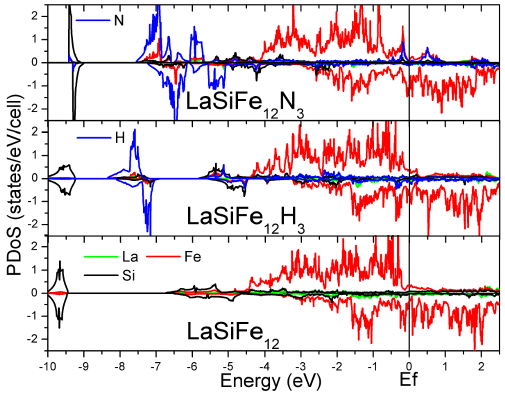}

\caption{\label{fig:PDOS} Partial density of states (PDOS) of the parent LaSiFe$_{12}$
alloy (bottom) in comparison with  LaSiFe$_{12}$H$_{3}$ (middle)
and LaSiFe$_{12}$N$_{3}$ (top). }
\end{figure}

Fig.~\ref{fig:PDOS} shows the PDOS of the parent alloy together with that of the fully hydrogenated and fully nitrogenated materials ($x=3$). In the parent alloy (bottom), the PDOS is dominated by Fe (red line) around the Fermi level (E$_{F}$) with typical spin-split states. The spin-up ($\uparrow$) states are mostly occupied, while the unoccupied states are dominated by spin-down ($\downarrow$) states separated by about 2.5eV in energy.  Furthermore, the filled bands at the lower end of energy range (-9.5~eV) relate mostly to silicon
3$s$ states which are overlapped with $p$-states of both La and Fe. A large energy gap appears from -9.5~eV up to about -6.5eV, where a high population of 3$p$ states of Si (black) is located (-6.5 to -4.5~eV). In this latter energy range, there is negligible contribution from Fe $d$-states. 

Most of the aforementioned features in the electronic structure are preserved in fully hydrogenated LaSiFe$_{12}$H$_{3}$ (middle of Fig.~\ref{fig:PDOS}). The main difference in the PDOS compared to
the parent alloy is the development of additional states in the gap around -7.5~eV related to the hydrogen interstitials. Small additional peaks also appear around -5~eV, where they overlap with the $p$ states of Si.  In strong contrast to hydrogenation, fully nitrogenated LaSiFe$_{12}$N$_{3}$ exhibits a large overlap of N $p$ states with Fe $d$ states in the -7.5 to -4~eV energy interval (top of Fig.~\ref{fig:PDOS}). These peaks indicate $p-d$ hybridization, and as a result, increased covalency in bond formation.   Such features also help to explain the non-monotonic moment as a function of doping (especially nitrogen).  Another important consequence of nitrogenation is the appearance of states in the vicinity of the Fermi level. 

The existence of a ``double peak'' feature just below and above E$_{F}$ for LaSiFe$_{12}$, created by Fe $d$ states in the minority DOS, is altered only a little by hydrogenation.  On the other hand, nitrogenation fills this valley at E$_{F}$, which results in the strong alteration of the magnetic properties and ultimately leads to the disappearance of IEM transition. We address the latter behavior in detail in the next section. 

\subsection{The free energy landscape }

We now turn our interest to the results of our second computational approach, fixed spin moment calculations using TB-LMTO.  Our aim is to visualize the energy difference between FM and NM states in the parent alloy and the doped materials. The purpose of our analysis is to identify
the main factors that lead to the field-induced isothermal entropy change of LaSiFe$_{12}$Z$_{3}$ around the magnetic transition being lower than that of LaSiFe$_{12}$, as found experimentally for interstitials other than hydrogen~\cite{chen_2003, teixeira_2012a, ZhangCdope, zhang_2010, balli_2009a,fujita_2003a}. 

\begin{figure}
\includegraphics[width=\columnwidth]{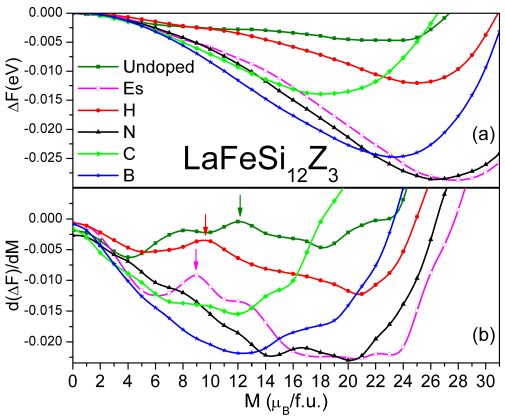}

\caption{\label{fig:dFvsM} Free energy of the FM state per atom (relative to the free energy of the non-magnetic state) as a function
of magnetization $M$, calculated by the fixed spin method (FSM) for LaSiFe$_{12}$ and for LaSiFe$_{12}$Z$_{3}$, ( Z = H, N, C, B, empty sphere (Es)). (b) The derivative, $d(\Delta F)/d M$.  Local minima in the free energy occur where $d(\Delta F)/d M=0$.}
\end{figure}

Fig.~\ref{fig:dFvsM} compares the free energy curves $F(M)$ calculated by the FSM method for LaSiFe$_{12}$Z$_{3}$, where
Z = H, N, B and empty sphere (Es), respectively, together with those of the
parent alloy.  For direct comparison, we set the non-magnetic energy
state as F(0) for each individual composition.  We also employ a constant volume approximation, which is used here to examine trends between differently doped compositions.  Fig.~\ref{fig:dFvsM}(a) shows that the parent LaSiFe$_{12}$ has a very shallow magnetic energy landscape, in accordance with  the predictions
by Kuz\textquoteright{}min and Richter, who used the full-potential local-orbital
(FPLO) method, also in a constant volume approach~\cite{Kuzmin_DFT}.  They noted the applied advantage of such a potential energy landscape in permitting a low hysteresis, first order metamagnetic transition (IEM).  The main difference in our calculation of the parent compound is that we find only two minima rather than the multiple minima that were predicted in their work.  The field-induced magnetisation of hydrogenated La-Fe-Si under pressure previously was seen to exhibit multiple steps, confirming Kuz\textquoteright{}min and Richter's predictions.  It may be interesting to investigate this property in the undoped compound to eliminate the possibility of  pressure-induced hydrogen segregation.  Spontaneous hydrogen segregation is known to occur in materials with a smaller hydrogen content than the empirical maximum~\cite{Barcza2011,Krautz2012,Zimm2013,Baumfeld2014}. 

The derivative of $\Delta F(M)$ with respect to $M$ is also shown in Fig.~\ref{fig:dFvsM}; local minima in the free energy function are where $d\Delta f/d M=0$. We may conclude that full doping of any of the four interstitials studied removes the shallow double-well potential, resulting in only a single well.  This corresponds to the disappearance of the first order metamagnetic transition, as found experimentally for B, C, and N doping.  We note that boron addition is detrimental to the total magnetisation as the minimum in $\Delta F(M)$ occurs at a lower value of $M$ than for any other dopant.  Of all the interstitials studied, hydrogen alters the double well picture the least.  The shallow landscape of $\Delta F(M)$ takes on a concave curvature in the range of $M\sim 7-9$~$\mu_{B}$ at full doping.    The intrinsically small energy barrier for the parent compound and the hydrogen-doped material makes these compositions particularly sensitive to external parameters such as magnetic field, pressure and temperature and renders the first order IEM transition quasi-reversible.

\begin{figure}
\includegraphics[width=\columnwidth]{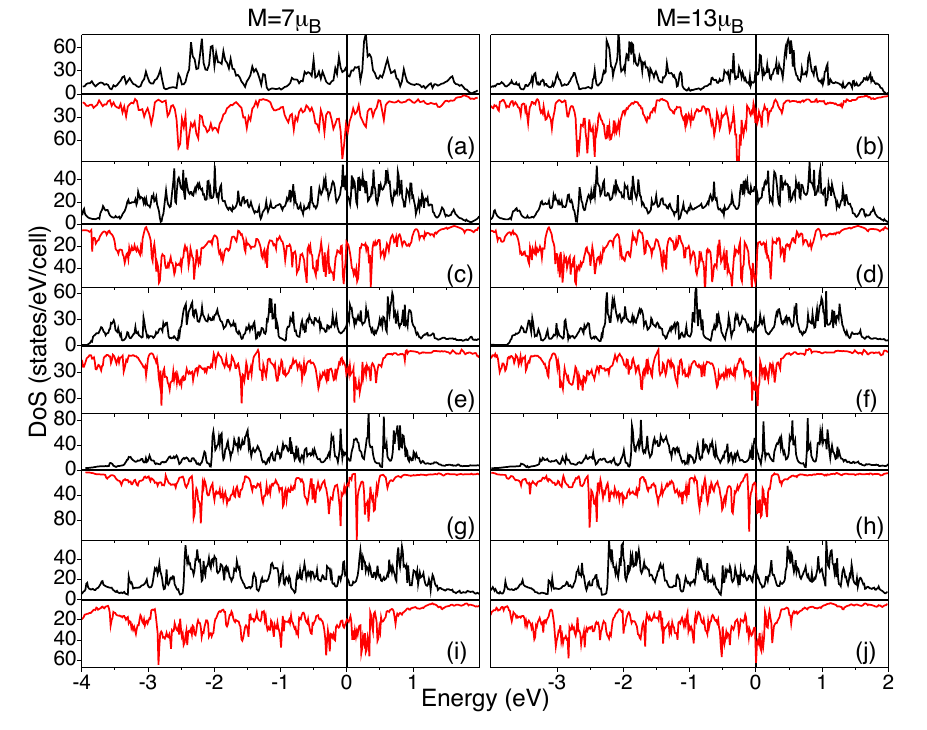}
\caption{\label{fig:FSMvsDOS} Up- and down DOS for LaSiFe$_{12}$ (i, j),
LaSiFe$_{12}$Es$_{3}$ (g, h), LaSiFe$_{12}$H$_{3}$ (e, f), LaSiFe$_{12}$N$_{3}$
(c, d) and LaSiFe$_{12}$B$_{3}$ (a, b) at M=7$\mu_{B}$ (left column)
and 13$\mu_{B}$ (right column) as calculated by FSM.}
\end{figure}

Fig. \ref{fig:dFvsM}b also reveals the sensitive nature of the metamagnetic states to the lattice parameters. An increase of about 0.5\% in the lattice parameter upon H-doping generates a ground state with a high-spin configuration.   In order to separate the changes in the magnetic state caused
by the volume expansion (chemical pressure) due to the inclusion of large interstitials such as nitrogen from those caused by additional valence electrons, we also carried out calculations with empty spheres (Es) included at the 24$d$ crystallographic site.  For a direct comparison, the same lattice constant was adopted for both Z = N and Es. 

To further examine the appearance of a metastable low-spin state at around $M=7\mu_{B}$ and the emergence of an energy barrier near $M=13\mu_{B}$, DOS calculations were performed with fixed spin moments at these two values of magnetisation for each system.  The results are shown in Fig. \ref{fig:FSMvsDOS}.  For both Z = H and Es, the Fermi level is located in a deep valley for both the minority and majority-spin DOS at M=7$\mu_{B}$, similar to that found
in the parent alloy. This well-defined valley in the DOS is destroyed for Z = B and N.  By contrast, the electronic features found in the DOS for M=13$\mu_{B}$ differ significantly: high peaks in the DOS appears at E$_{F}$ for both minority and majority-spin states for the undoped compound as well as for
Z=H and Es alloys, while only moderate peaks for Z = N and B are observed. It is thus apparent that the valleys
and peaks around E$_{F}$ can be attributed to the low-spin state and the energy barrier in $\Delta F(M)$. 

\begin{figure}
\begin{center}
\includegraphics[width=\columnwidth]{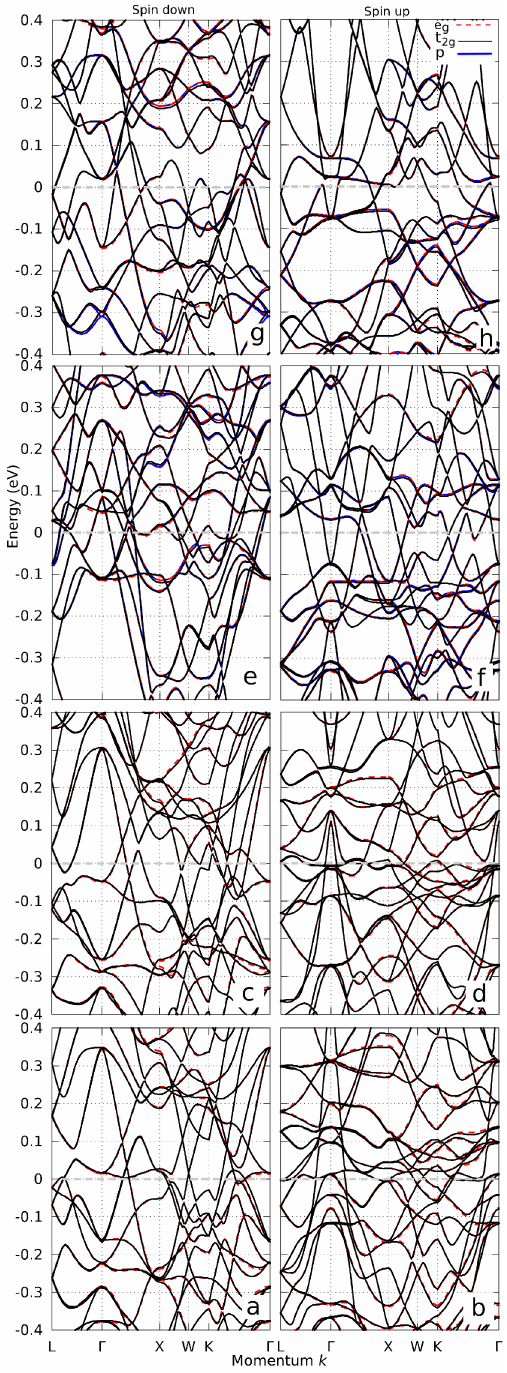}
\caption{\label{fig:FSMvsBdisp} Up- and down spin band dispersions for LaSiFe$_{12}$
(a,b), LaSiFe$_{12}$H$_{3}$ (c,d) and for LaSiFe$_{12}$N$_{3}$
(e,f) and LaSiFe$_{12}$B$_{3}$ (g,h) at M=13$\mu_{B}$ as calculated
by FSM. }  
\end{center}
\end{figure}

In Fig. \ref{fig:FSMvsBdisp}, we show the electronic band structure calculated for $M=13\mu_{B}$, in order to provide an explanation for the above peaks in the electronic DOS.  We note that a 0.01 eV offset was added to the energy level of p bands of the N-, B- and H-doped compounds for clarity in the figure.  We may make several qualitative observations. First, the quasi-degenerate t$_{2g}$ and e$_{g}$ bands are widely observed toward representative
k-directions along the Brillouin zone in the parent alloy. The additional
charge supplied by the H atoms shifts the position of the Fermi level
upwards but the character of these 3$d$ bands is conserved.  Second,
in the empty sphere configuration (Z = Es), where the significantly
larger lattice structure of the nitrogenated alloy is adopted without
any addition of charge, strong narrowing of the bandwidth is
found but the dispersion of each band is once again preserved (not shown).  

Third, the situation is very different for both Z=N and B, as the band structure is strongly altered at full doping. For Z=B, strong
$p-d$ band mixing occurs between -0.3 and -0.6~eV for the majority spin
states in all  $k$-directions. In addition, broad bands originating from the $p$-states of boron appear at the $\Gamma$ point about -0.3~eV
for the minority spins. The formation of these hybrid $p-d$ bands ultimately results in the vanishing of the well-formed peak
and valley structures of the DOS around E$_{F}$ as shown in Fig.~\ref{fig:FSMvsDOS} and in Fig.~\ref{fig:PDOS}.   Finally,
for nitrogen doping most of the $p$-states appear just above E$_{F}$, apart
from some minor ones around -0.5 eV in the W direction for the minority spins.  Here, the $p-d$ mixing occurs mainly between bands in the K and W directions.  Bands of $e_{g}$ character show especially pronounced mixing with $p$-electrons. The number of flat bands around E$_{F}$ is decreased
as compared to the undoped system and as a result the DOS has uneven features with small peaks that are detrimental to the IEM.
transition. 

\subsection{Electron coupling and inferred phonon entropy changes}
We may use the electronic DOS calculated in the previous sections to predict certain experimental quantities.  In this section we investigate the first of two: the electronic entropy change (and thereby the phonon entropy change) expected at the Curie temperature.  In the next section we will examine the variation of thermopower in the ferromagnetic state.   Both have been the subject of recent experimental work.

Several authors have attempted to decompose the entropy change at a first order magnetic phase transition into parts that can be ascribed to changes in the magnetic, phononic and electronic degrees of freedom.  This has especially been of interest in magnetocaloric materials studies, with examples including antiferromagnetic metamagnets such as CeFe$_{0.9}$Co$_{0.1}$~\cite{wada_1993}., (Fe$_{1-x}$Ni$_{x}$)$_{0.49}$Rh$_{0.51}$ with $x\sim0.03$~\cite{kreiner_1998}., and CoMnSi~\cite{barcza_2013} and the itinerant metamagnetic system considered in this article.   A significant point of divergence in approach occurs between adopting an itinerant or a localised view of the principle magnetic moments that order at the phase transition and also in the treatment of the phonons as Debye-like or otherwise.    Early work by Jia~\cite{jia_2006} proposed that upon ferromagnetic ordering, the entropy associated with a localised Fe moments decreased.  In La(Fe,Si)$_{13}$ compositions, there is a negative thermal expansion at the first order Curie point.  Jia et al. proposed that the material behaved as a phonon gas and that the larger ferromagnetic volume resulted in a counteracting, positive phonon entropy change.   However, a more recent inelastic resonant x-ray (INRXS) study by Gruner et al. found that the phonon entropy change on entering the ferromagnetic state was conventional (negative) despite the negative thermal expansion associated with the transition~\cite{gruner_2015}.

We have therefore set out to examine the first order phase transition from a partly itinerant electron standpoint.  The itineracy of the 3$d$ Fe electrons in La(Fe,Si)$_{13}$ is well established~\cite{fujita_2006a}.   There is also mounting evidence that the magnetism of La(Fe,Si)$_{13}$ is intermediate between a full itineracy and full localisation.  The presence of disordered local moments (DLMs) in the paramagnetic state has been supported by anomalous Hall effect measurements~\cite{fujita_2017a}, photoemission~\cite{kamakura_2010a}, coherent potential approximation (CPA) calculations~\cite{fujita_2012a, fujita_2016a} and recent fixed spin moment calculations~\cite{gruner_2017}.  Since the DLM moment may be of the order of 1~$\mu_{B}$, we calculate the field-induced change of magnetic entropy as a fixed value, $-R \textrm{ln}(2J+1)$, where we set $J=0.5$ rather than the value of $J=1$ that Jia et al. used, to reduce the possibility of overestimating the magnetic contribution (since the expression $\Delta S_{M}=-R \textrm{ln}(2J+1)$ yields an overestimate for the true change in magnetic entropy at a finite temperature, first order phase transition, even if the appropriate value of $J$ is used).  The actual value of $J$ or of $\Delta S_{M}$ is not of primary importance here; as we will see in Table~\ref{DeltaSTable1}, the trends between materials and between FM and PM states are retained by a constant shift in the value of $\Delta S_{M}$ chosen.  We calculate a ``bare'' electronic entropy as a function of temperature via the usual fermionic entropy relation:
\begin{eqnarray}
S_{\mathrm{el}}=-k_{B}&\int{g(E) [f \log f} \nonumber\\
& {+ (1-f)\log(1-f)]dE}.
\label{Eq:el-entropy}
\end{eqnarray}
where $k_B$ is the Boltzmann constant, $f$ is the fermi function and $g(E)$ is the density of states.  The chemical potential is adjusted self-consistently at different temperatures by fixing the total electron number.  We are thereby able to calculate bare electron Sommerfield coefficients, as the gradient of $(S_{el},T)$ curves where the non-magnetic (NM) state is taken as a proxy for the actual paramagnetic state.  We find that there is a near-linear relationship between electronic entropy and temperature in the non-magnetic state while it is almost perfectly linear in the FM state.   We note that the full Sommerfeld coefficient may be estimated on the basis of a modified free electron relation:
\begin{equation}
\gamma={1 \over 3} \pi^2  k_{B}^2 (1+\lambda)g(E_{F}) \, ,
\label{Eq:el-ph_coupling}
\end{equation}
where $\lambda$ is the magnitude of all other couplings of the electrons to phonons, spin fluctuations and so on and $E_{F}$ is the Fermi energy.  We may use this relation to estimate the strength of the electron-coupling mechanisms in the FM and PM state.

For LaSiFe$_{12}$, our estimate of the bare $\gamma^{FM}$ (27-29~mJK$^{-2}$kg$^{-1}$ in the FM state) depends on whether the density of states or gradient method is used (i.e. with $\lambda=0$) and is fairly close to theoretical estimates made by Gruner at al.~\cite{gruner_2015}.  However, all such estimates are a magnitude of about ten lower than experimental values such as those found by Fang et al.~\cite{fang_2008} (236~mJK$^{-2}$kg$^{-1}$ for LaAl$_{1.5}$Fe$_{11.5}$) and $\sim 3-4$~times lower than those found by Fujita et al.,~\cite{fujita_2003b} on La(Si$_{0.12}$Fe$_{0.88}$)$_{13}$ (LaSi$_{1.56}$Fe$_{11.44}$) and by Lovell et al.~\cite{lovell_2016}(~100-120 mJK$^{-2}$kg$^{-1}$ in samples of LaSi$_{1.2}$Fe$_{11.8}$ and LaSi$_{1.6}$Fe$_{11.4}$).   Since there is strong variation in the experimental value of $\gamma$ between those samples containing Si and those containing Al, we take the early data of Fujita et al.~\cite{fujita_2003b}, and the recent data on LaSi$_{1.2}$Fe$_{11.8}$ by Lovell et al. as our reference points for $\gamma^{FM}$ of our Al-free material models.  Both give $\gamma\sim 100$~mJK$^{-2}$kg$^{-1}$.

Information about the paramagnetic Sommerfeld coefficient is lacking and thus is predominately guided by theoretical, rather than empirical, considerations.   The Sommerfeld $\gamma^{PM}$ values that we find for the PM state of the hydrogen-doped LaSiFe$_{12}$H$_{3}$ compound vary somewhat, depending on whether the free electron formula in Equation~\ref{Eq:el-entropy} (yielding 60.7~mJK$^{-2}$kg$^{-1}$) or the gradient method (yielding 78.2 mJK$^{-2}$kg$^{-1}$) is used.  This can perhaps be attributed to the strong variation in the DOS in the immediate vicinity of the Fermi energy.  (The method-dependence of the Sommerfeld coefficient in the FM state of this compound and in the FM or PM states of LaSiFe$_{12}$ is considerably less.)

To gain more insight, we infer possible phonon contributions to the total isothermal entropy change, $\Delta S$, induced by a magnetic field, motivated by the contradiction between the recent observations by Gruner et al. and Jia and co-workers' early theoretical predictions.  The fact that the Sommerfeld $\gamma^{FM}$ coefficients are significantly lower than observed in low temperature experiments invites us to investigate equation (2) with $\lambda \neq 0$ i.e. with finite coupling between the electrons and phonons, spin fluctuations, and so on.  However, we will be unable to distinguish between such sources of mass enhancement of the electron.  We thus consider that $\Delta S$ is given principally by 
\begin{equation}
\Delta S = \Delta S_{M} + \Delta S_{el} + \Delta S_{el-coupling} + \Delta S_{ph} \, ,
\end{equation}
where the subscripts indicate the magnetic, bare electron, electron-coupling and pure phonon terms, respectively. From Equation~\ref{Eq:el-ph_coupling} the $\Delta S_{e-coupling}$ term above is given by $\lambda \times \Delta S_{el}$.  In Table~\ref{DeltaSTable1} we have compared the entropy changes that we find in LaSiFe$_{12}$ due to the bare electron and the coupled electron-phonon terms at $T_C = 195$~K.   The factor of 3.5 discrepancy between the $\gamma^{FM}$ values found here and those inferred from experimental data would imply that $\lambda^{FM}$ in the FM state is about 2.5, which is in a similar range to the value for spin fluctuation enhancement of gamma (3.3) found by Michor and co-workers in paramagnetic LaCo$_9$Si$_4$~\cite{michor_2004}.  To mimic the effect of magnetic field we calculate the difference in entropy between the FM and PM states as $\Delta S = S^{PM}-S^{FM}$.

\begin{table*}
\caption{\label{DeltaSTable1}
Calculated (DFT) and experimental~\cite{lovell_2016} Sommerfeld $\gamma$ factors and electronic entropy changes on cooling through the Curie transition of  LaSiFe$_{12}$ ($T_C=195$~K) and LaSiFe$_{12}$H$_3$ ($T_C$ extrapolated to 488~K).   All values are given in JK$^{-2}$kg$^{-1}$.  $\gamma_{el}$ values have been found by taking the gradient of $(S_{el},T)$ curves with respect to temperature.  $S_{el}$ values are obtained using Equation~\ref{Eq:el-entropy}.  The changes in entropy are negative because they are calculated by subtracting the entropy in the NM (PM) state from that of the FM state.  The values of $\lambda^{PM}$ and $\lambda^{FM}$ are allowed to vary.  For LaSiFe$_{12}$ we take $\lambda^{FM}=2.5$ while values for $\lambda^{PM}=2.5$ and $\lambda^{PM}=0.5$ are shown. For LaSiFe$_{12}$H$_3$ we take $\lambda^{FM}=1.5$ and  $\lambda^{PM}=0.5$ or $\lambda^{PM}=0.2$.  The (DLM) magnetic and electronic entropy changes are used to estimate a phonon entropy change, $\Delta S_{ph}$.  Our results show that consistency with the recent INRXS measurements of Gruner et al. requires that $\lambda^{PM}< \lambda^{FM}$ and that the effects of electron coupling in both magnetic states decreases with H-doping.}
\begin{indented}
\item[]\begin{tabular}{@{}llllllll}
\br
Material&$\gamma_{DFT}^{FM}$&$\gamma_{expt}^{FM}$~\cite{lovell_2016}&$\gamma_{DFT}^{PM}$&$S_{el}^{FM}$ & $S_{el-coupling}^{FM}$&$S_{el}^{PM}$&$S_{el-coupling}^{PM}$ \tabularnewline
\mr
LaSiFe$_{12}$&0.029&0.100& 0.075&5.6&14.0&16.6&41.5 \tabularnewline
LaSiFe$_{12}$&0.029&0.100& 0.075&5.6&14.0&16.7&8.4  \tabularnewline
LaSiFe$_{12}$H$_3$ &0.035&0.090& 0.078&17.2&25.8&39.6&19.8  \tabularnewline
LaSiFe$_{12}$H$_3$ &0.035&0.090& 0.078&17.2&25.8&39.6&7.9  \tabularnewline
\br
\end{tabular}
\item[]\begin{tabular}{@{}llcll}
\br
Material&$\Delta S_{M}$&$\Delta S_{el}+\Delta S_{el-coupling}$ & $\Delta S_{ph}$ (est.) & Conditions \tabularnewline
\mr
LaSiFe$_{12}$&-6.9&-38.5&+25.4& $\lambda^{FM}=2.5$; $\lambda^{PM}=2.5$ \tabularnewline
LaSiFe$_{12}$&-6.9&-5.5&-7.6& $\lambda^{FM}=2.5$; $\lambda^{PM}=0.5$  \tabularnewline
LaSiFe$_{12}$H$_3$ &-6.9&-16.4&+3.3& $\lambda^{FM}=1.5$; $\lambda^{PM}=0.5$  \tabularnewline
LaSiFe$_{12}$H$_3$ &-6.9&-4.5&-8.6& $\lambda^{FM}=1.5$; $\lambda^{PM}=0.2$  \tabularnewline
\br
\end{tabular}
\end{indented}
\end{table*}

If in the case of LaSiFe$_{12}$ we set $\lambda^{PM}=\lambda^{FM}=2.5$ so that $\gamma_{FM}$ is brought closer to the values found in experiment, we find that $\Delta S_{M}+\Delta S_{el} + \Delta S_{el-ph}=-45.4$~\JkgK and we are forced to infer a very large negative (positive) phonon entropy change on entering (leaving) the FM state since $\Delta S_{ph}=\Delta S-(\Delta S_{M}+\Delta S_{el} + \Delta S_{el-coupling})$ where $\Delta S \sim$ -20 \JkgK.   If however, we set $\lambda^{PM}<\lambda^{FM}$, then it is possible to adjust the inferred phonon contribution to a more reasonable range, in agreement with the findings of Gruner et al., as demonstrated in Table~\ref{DeltaSTable1}.  For $\lambda^{FM}=2.5$ and $\lambda^{PM}=0.5$  we find $\Delta S_{M}$+$\Delta S_{el} + \Delta S_{el-coupling}\sim-12.4$~\JkgK, and so $\Delta S_{phonon} \sim -7.6$~\JkgK. It should be noted that this level of suppression of $\lambda^{PM}=0.5$ is not the lowest possible (which is $\lambda^{PM}=0$); further suppression of $\lambda^{PM}$ would result in more negative changes in phonon entropy while higher $\lambda^{PM}$ values would (contrary to the recent experimental work by Gruner et al.) again infer a positive phonon entropy change.  The parameter set $\lambda^{FM}=2.5$ and $\lambda^{PM}=0.5$ delivers an approximately equal division of entropy change between electronic and purely phononic effects.

A similiar, if even more stark, conclusion can be drawn from an analysis of the fictitious LaSiFe$_{12}$H$_3$ compound.  The recent experimental data of Lovell et al. demonstrate that the low temperature,  $\gamma^{FM}$ values of La-Fe-Si compounds decrease with hydrogenation.  However, the bare $\gamma^{FM}$ value we find in DFT is slightly greater in the hydrogenated material.  Therefore, we must lower the $\lambda^{FM}$ value for the hydrogenated compound to around 1.5.  Since $\Delta S \sim$ -20 \JkgK in this family of materials, if we consider $\lambda^{PM}=0.5$, the resulting magnetic+electron entropy change is $\Delta S_{M} + \Delta S_{el} + \Delta S_{el-coupling}=~-23.3$~\JkgK (see Table~\ref{DeltaSTable1}).  Such a reduced value of $\lambda$ in the paramagnetic state yields only a very small phonon entropy change of around $\Delta S_{phonon} \sim +3.3$~\JkgK.  Given that this fictitious LaSiFe$_{12}$H$_3$ compound is basically second order (see earlier), our analysis implies that hydrogenation decreases the mass enhancement in both the FM and PM states, and reduces the role in the total entropy change played by the purely phononic term.  In Table~\ref{DeltaSTable1} we also show another parameter set for modelling the LaSiFe$_{12}$H$_3$ compound.  We may see that by lowering $\lambda^{PM}$ slightly, the balance of electronic and phonon entropy can be maintained at the level modelled in the case of LaSiFe$_{12}$.   The principal conclusion is therefore that the coupling of electrons to other degrees of freedom is reduced in the PM state and also by hydrogenation.  We have been able to achieve this perspective through the use of DFT-derived Sommerfeld $\gamma$ factors, and a comparison with experimental data available for the La(Si,Fe)$_{13}$H$_{\delta}$ material family.

The division of entropy, as noted above, is done with the strict assumption of there being purely a (DLM) magnetic, itinerant electronic, phonon, and electron-coupled term. However, there are other considerations, some of which our model has not considered.   Firstly, we have not considered the effects of spin waves at finite temperatures; we refer the reader to the recent analysis of spin waves from heat capacity data in Mn-containing La-Fe-Si compounds~\cite{lovell_2016}.  Further experiments are therefore warranted and will be important in determining a electronic coupling (and thereby the required phonon entropy change) more precisely as a function of hydrogenation.

\subsection{Thermopower}
We make a final comparison with experimental data using the electronic DOS calculated here.  Hannemann et al.~\cite{hannemann_2012} measured the thermopower in the ferromagnetic state of an unhydrogengated LaSi$_{1.4}$Fe$_{11.6}$ and hydrogenated  LaSi$_{1.4}$Fe$_{11.6}$H$_{\delta}$ and observed a broadening of the negative thermopower response in the latter material.  They ascribed this to a broadened electronic density of states, applying a paramagnetic (single spin) electron model used by Burkov et al.~\cite{burkov_1988} to the thermopower in the ferromagnetic state.

We may reverse the comparison by trying to predict the thermopower of LaSiFe$_{12}$ in the FM state while taking into account the effect of both spin channels.  We do so to motivate further work in this area as the level of agreement we find with available experimental data is mixed.  We employ a high temperature spin mixing approximation, such that the total thermopower is the average of that found in the two spin bands~\cite{macdonald_1962}.  We invoke one approximation that Hannemann et al. and Burkov et al. also used in their analysis; namely that the electronic conductivity is a separable function of temperature and of the energy of an electron and that it can be considered as inversely proportional to the electronic density of $d$-states.  Then the thermopower may be obtained from the linearised Boltzmann relation:

\begin{equation}
\mathcal{S}=-{1\over eT}{\int_{0}^{\infty}{\sigma(E,T)(E-E_{F})(-{\partial f \over \partial E}) dE}
\over
\int_{0}^{\infty}{(-{\partial f \over \partial E}) dE}
} \, .
\end{equation}

\begin{figure}
\includegraphics[width=\columnwidth]{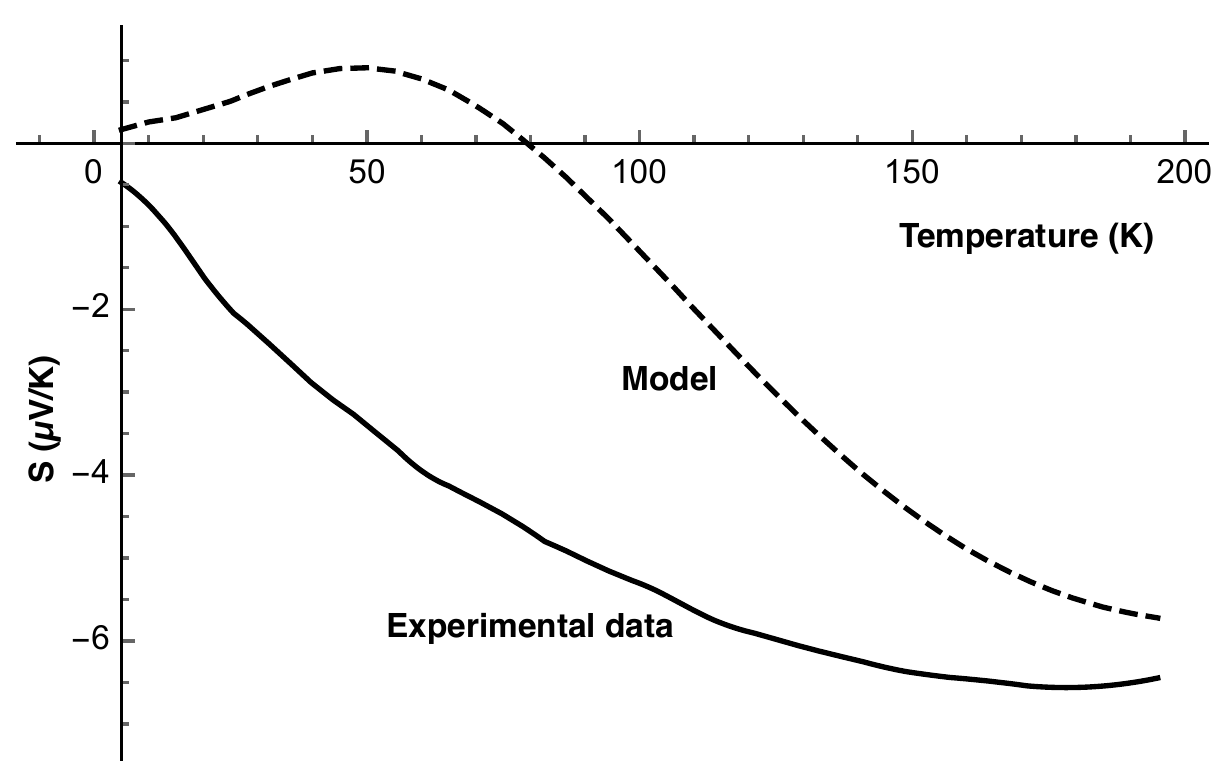}
\caption{\label{fig:thermopower} The calculated thermopower for LaFe$_{12}$Si (dotted line) compared with that obtained experimentally on LaSi$_{1.6}$Fe$_{11.4}$ by Hannemann et al.~\cite{hannemann_2012}. We note the similarity in magnitude and structure for the low temp regime, save for an additional feature at around 70~K in the modelled curve. }  
\end{figure}

We also allow essentially no variation of the magnetisation (chemical potential) within the FM state with temperature, up to the first order Curie point at 195~K.  For the non-hydrogenated compound, this is a reasonable approximation away from the Curie temperature.

Our results for the non-hydrogenated compound are shown in Figure~\ref{fig:thermopower}, along with experimental data from Hannemann et al. extracted from the best fit curve to data on a bulk sample of LaSi$_{1.4}$Fe$_{11.6}$.   We clearly see that some aspects of the thermopower in the FM state are reproduced by our simplified model. The broad form and magnitude are similar, but the sign of the low temperature thermopower is incorrect and the theoretical curve contains an additional feature at around 50 K which is due to band structure features but which is not seen in experiment.  Experimental thermopower data on the fictitious $x=3$ hydrogenated material are not available.  A comparison using a band structure calculation for $x=1.5$ published elsewhere~\cite{gercsi_2015} results in a poor agreement with experiment (not shown), even when temperature variation of the magnetisation (chemical potential) is included.  The calculated thermopower for $x=1.5$ increases from 100-300~K, in contrast to the experimental data.  We nonetheless note three aspects of our comparisons of calculated thermopower with experiment: (i) the relative success of thermopower modelling in the non-hydrogenated compound; (ii) the observation of theoretical features at low temperatures that can arise solely from band structure effects; and (iii) that scattering effects beyond our model are likely to play a larger role in the accurate modelling of the thermopower in the hydrogenated La-Fe-Si samples.  It has already been suggested that experimental features in the low temperature thermopower of the hydrogenated material are due to the presence of such effects~\cite{hannemann_2012,lovell_2016}.  Further work will be required to separate the effect of band structure from additional scattering mechanisms.

\section{Conclusions}
\label{sec_conc}

We have investigated the effect of select $s$- and $p$-block interstitial elements on the electronic, lattice and magnetic properties of
La(Fe,Si)$_{13}$ using DFT.   Our calculations find that a good correlation between the size of the unit cell and the size of the dopant.  Fixed spin moment calculations yield a double well structure in the free energy of LaFe$_{12}$Si as a function of magnetisation, which may be seen as the basis of the itinerant electron metamagnetic transition.  Significantly, hydrogenation alters the electronic and magnetic structure
of LaSiFe$_{12}$ to a much smaller degree than B, C and N dopants.  This means that the first order IEM is much more robust to H insertion than to interstitial B, C, or N.  

An analysis of the projected electronic DOS reveals that the dominant electronic states related to hydrogen insertion appear at around -8 to -7~eV, where very little contribution from Fe, Si and La elements is present. The additional charge of the hydrogen atoms elevates the Fermi level but the character of the bands is unaltered. The latter feature is also evident in the empty sphere configuration, where the nitrogenated lattice parameters are simulated without the  N inclusions; only a narrowing of the band-width is found but the dispersion of each band remains mostly unaffected.

Consequently, hydrogen provides perhaps the only chemical pressure on the lattice that avoids significant alterations to the electronic structure of LaSiFe$_{12}$. In the case of the other dopants (B, C, N), broad bands originating from their $p$-states appear at energy levels where the 3$d$ states of Fe are also present. The formation of these hybrid $p-d$ bands results in the disappearance of the peak and valley structures in the electronic DOS around E$_{F}$, thereby reshaping the shallow free energy landscape and ultimately destroying the first order IEM of LaSiFe$_{12}$.   Our theoretical findings are in good agreement with the experimentally-determined properties of interstitially doped La(Si,Fe)$_{13}$ compounds~\cite{chen_2003,teixeira_2012a,ZhangCdope,zhang_2010,balli_2009a,fujita_2003a}.

We have used our calculated band structure for the non-hydrogenated and hydrogenated materials to predict a range of Sommerfeld $\gamma$ enhancements that would achieve the same sign of phonon entropy change at the magnetic ordering transition that Gruner et al. have reported.  We demonstrate that the $\gamma$ enhancement of the FM state is greater than that of the PM state, consistent with a picture that such an enhancement comes mainly from spin fluctuations.   Hydrogenation lowers the relative enhancement of the electronic heat capacity (implying a reduction of spin fluctuation effects) and the bare electronic $\gamma$  is slightly increased.  We have further calculated thermopower, in good agreement with experiments on non-hydrogenated material.   Thermopower values in the hydrogenated material seem to require a model that includes other scattering effects, although these are unlikely to spin waves, which are thought to be heavily suppressed~\cite{lovell_2016}.

\ack
The authors thank L.F. Cohen for useful discussions. The research leading to these results has received funding from the European Community's 7th Framework Programme under Grant Agreement No. 310748 "DRREAM". Computing resources provided by Darwin HPC and Camgrid facilities at The University of Cambridge and the HPC Service at Imperial College London are also gratefully acknowledged.  NF acknowledges the Brooklyn College LSAMP program and the NSF for funding. This material is based upon work supported by the National Science Foundation under Grant no. 1202520 (NYC Louis Stokes Alliance).

\section*{References}
\bibliography{LaFeSiYx_JPhysD-format_v3_arxiv}

\end{document}